\documentclass[aps,prb,reprint,amsmath,amssymb]{revtex4-2}
\usepackage[bookmarks=false,linkcolor=NavyBlue,urlcolor=NavyBlue,colorlinks,citecolor=NavyBlue]{hyperref}
\usepackage{graphicx}
\usepackage{amsmath}
\usepackage{bm}
\usepackage[svgnames]{xcolor}
\usepackage{enumerate}
\usepackage[normalem]{ulem}
\usepackage{microtype}
\usepackage{comment}

\renewcommand{\vec}[1]{\mathbf{#1}}

\newcommand{\U}{{\rm U}}
\newcommand{\UA}[1]{#1_{\uparrow}}
\newcommand{\BDA}[1]{\overline{#1}_{\downarrow}}

\newcommand{\Cornell}{Department of Physics, Cornell University, Ithaca, NY 14853, USA}

\newcounter{para}
\newcommand{\para}{\par\refstepcounter{para}\textsf{ \textbf{{\color{Purple}[\thepara]}}\space}}
\let\para\relax

\begin{document}
\title{Trial wavefunction for fractional quantum spin Hall insulators}

\author{Omri Lesser}
\affiliation{\Cornell}

\author{Chao-Ming Jian}
\affiliation{\Cornell}

\begin{abstract}
Fermions with opposite spins occupying half-filled conjugate Chern bands exhibit interaction physics distinct from their multi-component Landau-level counterparts with the same chirality. This is largely due to unavoidable inter-species collisions that preclude the Halperin-type wavefunctions available in multi-component Landau levels. In this work, we propose and evaluate a variational wavefunction for a fractional quantum spin Hall state with $\mathbb{Z}_4$ topological order in a pair of conjugate Landau levels. This $\mathbb{Z}_4$ topological order has previously been shown to be the minimal topological order compatible with charge conservation, $S_z$ conservation, time-reversal symmetry, and the fractional spin Hall conductance $\sigma^{\rm sh}=\frac{1}{2}$ suggested by previous twisted $\text{MoTe}_2$ experiments. Our construction is based on the condensation of an anyonic exciton formed by the neutral fermionic excitations in a decoupled pair of Moore-Read Pfaffian state and its conjugate. By coupling the chiral and anti-chiral Ising conformal field theories associated with the two spin species, we introduce a variational mass parameter in the $\mathbb{Z}_4$ trial wavefunction that captures the inter-spin-species $s$-wave pairing of composite fermions alongside the intra-spin-species $p$-wave pairing. We assess the energetics of this trial state using Monte Carlo sampling on a spherical geometry. Because the coupled state intrinsically involves Landau-level mixing, we explicitly evaluate the resulting kinetic energy penalty. Our phase diagram reveals that the proposed $\mathbb{Z}_4$ state becomes energetically favorable in a sizable region of parameter space, over both the decoupled pair of conjugate Pfaffian states and an alternative exciton condensate state. These results provide a concrete microscopic wavefunction realization of this $\mathbb{Z}_4$ fractional quantum spin Hall phase, and propose a route to constructing additional families of such states.
\end{abstract}
\maketitle

\section{Introduction}

\para
The realization of fractional topological insulators and fractional quantum spin Hall states has long been a subject of profound theoretical interest, generalizing the physics of fractional quantum Hall effects to systems with time-reversal (TR) symmetry and conjugate Chern bands~\cite{levin_fractional_2009,regnault_fractional_2011,repellin_mathbbz_2_2014,neupert_fractional_2011,neupert_fractional_2015,stern_fractional_2016}. While initially confined to theoretical proposals, fractional states in partially filled Chern bands have recently become experimentally relevant thanks to rapid advances in the fabrication of moir\'{e} materials. Recent seminal experiments have reported evidence of these exotic states in twisted homobilayer transition metal dichalcogenides (TMDs), particularly twisted $\text{MoTe}_2$~\cite{kang_evidence_2024,kang_time-reversal_2025,wang_magnetic_2026}. Notably, transport measurements in twisted $\text{MoTe}_2$ have found evidence that suggests an incompressible state in half-filled conjugate Chern bands, with a fractional spin Hall conductance $\sigma^{\rm sh}=\frac{1}{2}$ (in units where a fully filled pair of conjugate bands contributes one unit)~\cite{kang_evidence_2024}. Furthermore, rich spin Hall physics has been observed in other highly tunable two-dimensional platforms, including twisted bilayer $\text{WSe}_2$~\cite{han_quantum_2025} and electric-field-tuned $\text{MoTe}_2/\text{WSe}_2$ heterobilayers~\cite{nguyen_quantum_2025,chang_electric-field-tuned_2026}. Together, these experimental breakthroughs provide a fertile ground for exploring two-component fractional phases in conjugate bands, necessitating a deeper theoretical understanding of their ground-state wavefunctions.
Since these platforms conserve both the electric charge and the (pseudo) spin component $S_z$, carried by the valley or layer degree of freedom, the resulting $\U(1)_c \times \U(1)_s$ symmetry, together with the TR symmetry of the system, poses significant constraints on candidate states.

\para
Fermions occupying conjugate Chern bands ($C = \pm 1$) exhibit fundamentally different interaction physics compared to their multi-component Landau level (LL) counterparts with the same chirality. In the latter case, where the bands carry equal Chern numbers, an inter-species Jastrow factor minimizes the contact-interaction energy: the factor $\prod_{i,j}(z_i - w_j)^{n}$ of a Halperin $(lmn)$ state (with $n \ge 1$) forces the wavefunction to vanish whenever two fermions of different species coincide, eliminating the energy cost of the short-range repulsion altogether~\cite{halperin_theory_1983,haldane_fractional_1983}. The opposing chiralities of the cyclotron motions in conjugate bands rule out this route: one species is described by holomorphic coordinates and the other by anti-holomorphic ones, so no analytic inter-species Jastrow factor exists, and it is challenging to write a wavefunction in which the fermions fully avoid one another~\cite{wagner_variational_2025}.
Inter-species collisions then lead to an energy penalty, suggesting an inherent instability toward the spontaneous breaking of time-reversal or translational symmetry, posing the central challenge that any candidate wavefunction in conjugate bands must confront.

\para
To address the unique kinematics of conjugate Chern bands, several alternative trial wavefunctions have been proposed. One candidate wavefunction at half-filling is a non-Abelian spin Hall insulator~\cite{abouelkomsan_non-abelian_2025} that operates as two conjugate copies of the Moore-Read Pfaffian state~\cite{moore_nonabelions_1991}: in each copy, composite fermions---electrons bound to two vortices---form a chiral $p$-wave paired state. By construction, however, this wavefunction describes the decoupled limit, with no inter-species correlations to mitigate the collision energy. While this non-Abelian spin Hall insulator phase can be stable within a regime of finite inter-species correlation~\cite{abouelkomsan_non-abelian_2025}, the candidate wavefunction becomes intractable (beyond exact diagonalization).
Another framework utilizes particle-hole Halperin states, such as the (331) state, which explicitly break time-reversal symmetry while maintaining zero Hall conductivity~\cite{sodemann_villadiego_halperin_2024}. These states notably feature disorder-resilient itinerant quasiparticles that may explain the lack of robust Hall plateaus in some experimental observations.
Motivated by the difficulty in stabilizing such Halperin-type states with repulsive interactions, a recent work has constructed a new class of composite fermion pairing wavefunctions~\cite{wagner_variational_2025}, in which composite fermions of opposite spins form $s$-wave pairs. This pairing strongly couples the two species, but the resulting state---a composite-fermion superconductor, or an exciton insulator when one species consists of holes---carries no fractional topological order.
In addition, an exact diagonalization study has investigated the possibility of fractional quantum spin Hall states in twisted $\text{MoTe}_2$~\cite{kwan_regarding_2026}.
A trial wavefunction for a candidate state that strongly correlates the two species while preserving all symmetries and supporting fractional topological order has been missing so far. Searching for such trial wavefunctions can deepen our understanding of the landscape of fractional incompressible states in conjugate LLs, especially in regimes where system sizes exceed the limits of exact diagonalization.

\para
Despite the lack of a wavefunction description, such a candidate topological state has recently been identified~\cite{jian_minimal_2025,zhang_non-abelian_2024}: a fractional quantum spin Hall (FQSH) state with $\mathbb{Z}_4$ topological order, which is the minimal topological order compatible with charge conservation, $S_z$ conservation, time-reversal symmetry, and the reported fractional spin Hall conductance $\sigma^{\rm sh}=\frac{1}{2}$~\cite{jian_minimal_2025,kang_evidence_2024}. This $\mathbb{Z}_4$ FQSH state is connected to the decoupled limit by the condensation of an ``anyonic exciton'': a self-bosonic bound state of the neutral fermionic excitations that any incompressible state in a half-filled LL supports~\cite{halperin_theory_1993,jian_minimal_2025}. The condensation generates strong inter-species correlations without breaking any symmetry. So far, however, the $\mathbb{Z}_4$ FQSH state has been constructed only through its minimal topological field theory and through various parton constructions, as a strong-pairing descendant of a vortex spin liquid~\cite{jian_minimal_2025, zhang_vortex_2024,zhang_non-abelian_2024}; neither construction is controlled in terms of energetics, leaving open what microscopic Hamiltonian can potentially stabilize this state.

\para
In this work, we construct and assess a variational wavefunction for the $\mathbb{Z}_4$ FQSH state in a pair of half-filled conjugate LLs filled with fermions with two spin species (viewed as a representative of conjugate Chern bands with ideal quantum geometry). We start in the decoupled limit from a Moore-Read Pfaffian wavefunction and its conjugate which admits a conformal field theory (CFT) description. We couple their associated chiral and anti-chiral Ising CFTs through a single variational mass parameter that captures the condensation of the anyonic exciton. In the resulting trial wavefunction, the composite fermions develop inter-spin-species $s$-wave pairing alongside the intra-spin-species $p$-wave pairing, yielding the $\mathbb{Z}_4$ FQSH topological order. Building on this construction, we provide recipes to construct wavefunctions with anyonic excitation allowed by its $\mathbb{Z}_4$ topological order. We evaluate the energetics of this trial state by Monte Carlo sampling in a spherical geometry, explicitly accounting for both the interaction energy and the kinetic-energy penalty of the LL mixing intrinsic to the coupled state. The resulting phase diagram shows that the $\mathbb{Z}_4$ FQSH trial wavefunction becomes energetically favorable over a sizable region of parameter space, outcompeting both the decoupled Pfaffian pair and the composite-fermion $s$-wave superconductor~\cite{wagner_variational_2025}. This region involves attractive inter-spin-species interaction, which can naturally arise when the fermions of the two pseudospins are electrons and holes respectively, a scenario that can occur in TMD heterobilayers~\cite{nguyen_quantum_2025,chang_electric-field-tuned_2026}.

\section{$\mathbb{Z}_4$ fractional quantum spin Hall wavefunction}

\para
The system we focus on in this work consists of fermions, $f_{\uparrow}$ and $f_{\downarrow}$, of two (pseudo) spin species. The single-particle physics of $f_{\uparrow}$ and $f_{\downarrow}$ is described by a pair of conjugate Landau levels or Chern bands with opposite Chern numbers. In twisted homobilayer $\text{MoTe}_2$~\cite{kang_evidence_2024,kang_time-reversal_2025}, the two species of fermions are electrons from the $K/K'$ valleys, whereas in the TMD heterobilayer context~\cite{nguyen_quantum_2025,chang_electric-field-tuned_2026} the two species are electrons and holes from the two different layers. In these systems, it is natural to consider the combination of charge $\U(1)_c$ symmetry, spin-$S_z$ symmetry $\U(1)_s$, and time-reversal symmetry. 

\para
We are interested in studying candidates of incompressible fractional states in a pair of half-filled conjugate LLs using a variational wavefunction approach. Such conjugate LLs are treated as an example of conjugate Chern bands with ideal quantum geometry~\cite{CanoWang2026}. For a single isolated half-filled LL, well-known candidates for incompressible states include the Moore-Read Pfaffian and anti-Pfaffian states~\cite{moore_nonabelions_1991,greiter_paired_1991,lee_particle-hole_2007,storni_fractional_2010}, the particle-hole Pfaffian~\cite{son_is_2015,rezayi_stability_2021}, the Abelian $\U(1)_{8}$ state~\cite{greiter_paired_1992}, or the Halperin (331) state~\cite{wen_classification_1992,overbosch_phase_2008}. Hence, a simple class of incompressible candidate states for the half-filled conjugate LLs, especially when the two spin species are decoupled, is given by conjugate pairs of the half-filled-LL states. These candidates respect the full $\U(1)_c\times \U(1)_s$ and TR symmetries. They all feature a fractional quantum spin-Hall conductance $\sigma^{\rm sh} = \frac{1}{2}$, consistent with the transport evidence in twisted $\text{MoTe}_2$ reported in Ref.~\cite{kang_evidence_2024}. 

\para
Refs.~\cite{jian_minimal_2025, zhang_non-abelian_2024} identified a new class of $\U(1)_c\times \U(1)_s$-symmetric FQSH candidates for the half-filled conjugate LLs with strong inter-spin correlations. These states also exhibit $\sigma^{\rm sh} = \frac{1}{2}$. Among them, there is a unique TR-symmetric candidate with a $\mathbb{Z}_4$ topological order. The previous studies of this state are based on topological quantum field theories and parton construction, which are energetically uncontrolled. In the following, we provide a complementary perspective on the $\mathbb{Z}_4$ FQSH state using a variational trial wavefunction that gives direct access to its energetics. We will focus on constructing and evaluating the wavefunction in a conjugate pair of lowest Landau levels. The wavefunction construction can be naturally generalized to conjugate Chern bands with ideal quantum geometry~\cite{CanoWang2026}.

\para
Our construction of the trial wavefunction for $\mathbb{Z}_4$ FQSH state is based on anyon condensation in a conjugate pair of decoupled half-filled LL states. A key observation is that any incompressible topological state in a half-filled LL supports a neutral fermionic excitation~\cite{halperin_theory_1993}.
If, in the decoupled limit, the conjugate LLs of the two spin species are described by an incompressible topological state in one LL and its conjugate in the other, then the neutral fermionic excitations in both states can bind to form an ``anyonic exciton".
Since this anyonic exciton has self-bosonic statistics, it can condense, resulting in a new state in which the two spin species are strongly coupled. One could choose different half-filled LL states and their conjugates in the decoupled limit of this construction. Interestingly, the resulting state is always topologically equivalent to the same $\mathbb{Z}_4$ FQSH state~\cite{jian_minimal_2025}. This construction motivates our trial wavefunction for the $\mathbb{Z}_4$ FQSH state, which starts from the non-Abelian Moore-Read Pfaffian state and its conjugate in the decoupled limit.

\para
Let us first briefly review the Moore-Read Pfaffian wavefunction and its CFT construction~\cite{moore_nonabelions_1991}. 
For the half-filled LL, this wavefunction takes the form 
\begin{equation}\label{eq:moore_read_pfaffian}
\begin{aligned}
    \Psi_{\rm MR} (\left\{z_{i}\right\}) &= {\rm Pf}\left(\frac{1}{z_{i}-z_{j}}\right) \prod_{i<j}(z_{i}-z_{j})^2\\
    &\times\exp\left(-\frac{1}{4\ell_{B}^2}\sum_i |z_i|^2\right),
\end{aligned}
\end{equation}
where $z=x+i  y$ is the complex holomorphic coordinate of an electron and $\ell_{B}$ is the magnetic length. This state describes an incompressible and non-Abelian $p_{x}+ip_{y}$ superconductor of composite fermions in a half-filled LL.
There, the $\prod_{i<j}(z_{i}-z_{j})^{2}$ Jastrow factor attaches two flux quanta (vortices) to each electron to form composite fermions.
The Pfaffian factor ${\rm Pf}\left(\frac{1}{z_{i}-z_{j}}\right)$ describes the pairing of the composite fermions. 

\para
The Moore-Read Pfaffian wavefunction can also be naturally written as a correlator of a 2D CFT given by the product of the $\U(1)_{8}$ chiral boson CFT and the chiral Ising CFT. The primary fields of this CFT are given by the products of the vertex operators $\left\{ e^{i k \phi},\ k=0,1,\ldots,7 \right\}$ in the $\U(1)_{8}$ chiral boson sector and the three primary fields $\left\{ 1,\sigma,\psi \right\}$ of the chiral Ising sector. Here, $\phi$ denotes a chiral boson field in $\U(1)_{8}$. $\psi$ is the chiral energy operator of the Ising CFT, which also represents a chiral Majorana fermion. The local electron operator is identified with $f\sim\psi e^{4i\phi}$. The Moore-Read Pfaffian wavefunction can be obtained as the correlator of the fermion operators
\begin{equation}
    \Psi_{\rm MR}(\left\{z_{i}\right\}) = \left\langle f(\vec{r}_1) f(\vec{r}_2) \cdots \right\rangle
\end{equation}
in this chiral ${\rm Ising} \times \U(1)_{8} $ CFT. Here, $\vec{r}_i=(x_i,y_i)$ denotes the vector form of the electron coordinate. This correlator factorizes into two terms: $\left\langle \psi(\vec{r}_{1})\psi(\vec{r}_{2}) \cdots \right\rangle_{\rm Ising}$ from the Ising sector gives rise to
the Pfaffian piece and $\left\langle e^{4i\phi(\vec{r}_1)} e^{4i\phi(\vec{r}_2)} \cdots \right\rangle_{\U(1)_{8}}$ yields the Jastrow factor and the Gaussian piece in Eq.~\eqref{eq:moore_read_pfaffian}. Since both the Ising and $\U(1)_{8}$ CFTs are chiral, these correlators only depend on the holomorphic coordinates. Wavefunctions with anyonic excitations on top of the Pfaffian state can be obtained by inserting the chiral primary fields $e^{i n \phi}$, $\psi e^{i n \phi}$, $\sigma e^{i n' \phi}$ with even $n \in 2\mathbb{Z}$ and odd $n'\in 2\mathbb{Z}+1$ into the correlator. More specifically, $e^{i n \phi}$ generates a charge-$\frac{n}{4}e$ Abelian anyon, $\psi$ generates the neutral fermionic excitation, and $\sigma e^{i n' \phi}$'s are associated with the charge-$\frac{n'}{4}e$ non-Abelian anyons.

\para
We now proceed to describe two conjugate copies of the Pfaffian state in a pair of conjugate half-filled LLs. The fermions in the two conjugate LL are distinguished by their (pseudo) spin indices $\uparrow$ and $\downarrow$.
The total Pfaffian$\times \overline{\rm Pfaffian}$ state has a wavefunction given by the correlator in a pair of chiral and anti-chiral CFTs, ${\rm Ising} \times \U(1)_{8} \times \overline{\rm Ising} \times \overline{\U(1)_{8}}$. Here, $\overline{~\cdot~}$ represents the ``anti-chiral-ness" (or anti-holomorphicity) of the CFTs and the associated states. We denote the chiral and anti-chiral Majorana fermion fields of the respective Ising and $\overline{\rm Ising}$ CFTs as $\psi_{\uparrow}$ and $\bar{\psi}_{\downarrow}$. $\phi_{\uparrow}$ and $\bar{\phi}_{\downarrow}$ denote the chiral and anti-chiral bosonic fields associated with the $\U(1)_8 \times \overline{\U(1)_8}$ charge sectors. The local fermion operators of two spin species are then represented as $f_{\uparrow}\sim\psi_{\uparrow}e^{4 i \phi_{\uparrow}}$ and $f_{\downarrow}\sim\bar{\psi}_{\downarrow}e^{4 i \bar{\phi}_{\downarrow}}$.

\para
The $\mathbb{Z}_4$ FQSH state can be obtained from condensing the anyonic exciton $\psi_{\uparrow}
\bar{\psi}_{\downarrow}$ in the Pfaffian $\times \overline{\rm Pfaffian}$ state~\cite{jian_minimal_2025}. At the level of a trial wavefunction, this condensation can be captured by coupling the two chiral and anti-chiral Ising CFTs. The coupled theory can be minimally described by the Lagrangian
\begin{equation}\label{eq:L_coupled_Ising}
    {\cal L} = \psi_{\uparrow} \partial_{\bar{z}} \psi_{\uparrow} + \bar{\psi}_{\downarrow} \partial_{z} \bar{\psi}_{\downarrow} + i m \psi_{\uparrow}\bar{\psi}_{\downarrow},
\end{equation}
where $m$ is a ``mass" term that captures the condensation of the anyonic exciton $\psi_{\uparrow}\bar{\psi}_{\downarrow}$.
We will treat it as a variational parameter describing the strength of the coupling between the spin species. The sector $\U(1)_8 \times \overline{\U(1)_8}$ of chiral and anti-chiral bosons will remain unchanged. The trial wavefunction for the $\mathbb{Z}_4$ FQSH state is then given by the correlator $\Psi_{\mathbb{Z}_{4}} =\left\langle f_\uparrow(\vec{r}_1) f_\uparrow(\vec{r}_2) \cdots f_\uparrow(\vec{r}_N) f_\downarrow(\vec{r}_1') f_\downarrow(\vec{r}_2') \cdots f_\downarrow(\vec{r}'_N) \right\rangle$ where $\vec{r}_i$ ($\vec{r}_i'$) is the vector form of the coordinate of a spin-up (spin-down) local fermion.  $N$ is the number of fermions of each spin.

\para
To evaluate this trial wavefunction, we start with the two-point function of the coupled Ising theory Eq.~\eqref{eq:L_coupled_Ising}:
\begin{equation}
\begin{aligned}   
& \begin{pmatrix}\left\langle \psi_{\uparrow}\left(\vec{r}\right)\psi_{\uparrow}\left(0\right)\right\rangle  & \left\langle \psi_{\uparrow}\left(\vec{r}\right)\bar{\psi}_{\downarrow}\left(0\right)\right\rangle \\
\left\langle \bar{\psi}_{\downarrow}\left(\vec{r}\right)\psi_{\uparrow}\left(0\right)\right\rangle  & \left\langle \bar{\psi}_{\downarrow}\left(\vec{r}\right)\bar{\psi}_{\downarrow}\left(0\right)\right\rangle 
\end{pmatrix}  \sim \begin{pmatrix}-\partial_{z} & -im\\
im & -\partial_{\bar{z}}
\end{pmatrix} G_{b} (\vec{r}),
\end{aligned}
\end{equation}
where the Green's function is 
\begin{equation}
G_{b} (\vec{r}) = \int \frac{d^2 \vec{p}}{(2\pi)^2} \frac{e^{i\vec{p}\cdot\vec{r}}}{\vec{p}^{2}+m^{2}} = \frac{i}{2\pi} K_{0} (m|\vec{r}|),
\end{equation}
with $K_{0}$ being the modified Bessel function of the second kind. Using the two-point correlators of the coupled Ising theory, we now write the corresponding trial wavefunction. Because the Lagrangian in Eq.~\eqref{eq:L_coupled_Ising} is quadratic in the fermion fields, the Ising sector of the $2N$-point correlator $\left\langle f_\uparrow(\vec{r}_1) f_\uparrow(\vec{r}_2) \cdots f_\uparrow(\vec{r}_N) f_\downarrow(\vec{r}_1') f_\downarrow(\vec{r}_2') \cdots f_\downarrow(\vec{r}'_N) \right\rangle$ can be evaluated using Wick's theorem.
This expectation value evaluates exactly to the Pfaffian of the $2N \times 2N$ covariance matrix constructed from the two-point Green's functions.
The chiral boson $\U(1)_{8}$ sector remains the same as in $\Psi_{\rm MR}$, while the $\overline{\U(1)_{8}}$ sector behaves similarly but in an anti-holomorphic manner. This leaves us with the full trial wavefunction of the $\mathbb{Z}_4$ FQSH state:
\begin{equation}
\begin{aligned}    
\Psi_{\mathbb{Z}_{4}}&\left( \left\{ u_{i} \right\}, \left\{ w_{i} \right\} \right) = \prod_{i<j}(u_{i}-u_{j})^{2} \prod_{i<j}(\bar{w}_{i}-\bar{w}_{j})^{2} \\
&\times \exp\left[-\frac{1}{4\ell_{B}^2}\sum_i \left(|u_i|^2 + |w_i|^2 \right) \right]
 {\rm Pf}(A).
\end{aligned}
\end{equation}
Here $\left\{ u_{i} \right\}$ and $\left\{ w_{i} \right\}$ are the complex holomorphic coordinates of the fermions with $\uparrow$ and $\downarrow$ spins (recall that their vector forms are denoted as $\vec{r}_i$ and $\vec{r}_j'$, respectively). With $N$ fermions per spin, $A$ is a matrix of two-point correlators in the coupled Ising theory Eq.~\eqref{eq:L_coupled_Ising}, which is composed of four $N\times N$ blocks, 
\begin{equation}\label{eq:A_blocks}
A = \begin{pmatrix}
A_{\uparrow\uparrow} & A_{\uparrow\downarrow} \\
A_{\downarrow\uparrow} & A_{\downarrow\downarrow}
\end{pmatrix},
\end{equation}
with the matrix elements
\begin{subequations}\label{eq:A_matrix_elements}
\begin{equation}
\begin{aligned}
(A_{\uparrow\uparrow})_{ij} &=\left\langle \psi_{\uparrow}\left(\vec{r}_i\right)\psi_{\uparrow}\left(\vec{r}_j\right)\right\rangle\\ &= -\frac{\partial K_0(m|r|)}{\partial z}\Big|_{r=u_i-u_j} \quad (i \neq j),
\end{aligned}
\end{equation}
\begin{equation} 
\begin{aligned}
(A_{\downarrow\downarrow})_{ij} &=\left\langle \bar{\psi}_{\downarrow}\left(\vec{r}_i'\right)\bar{\psi}_{\downarrow}\left(\vec{r}_j'\right)\right\rangle\\ &= -\frac{\partial K_0(m|r|)}{\partial \bar{z}}\Big|_{r=w_i-w_j} \quad (i \neq j),
\end{aligned}
\end{equation}
\begin{equation} 
\begin{aligned}
(A_{\uparrow\downarrow})_{ij} &= \left\langle \psi_{\uparrow}\left(\vec{r}_i\right)\bar{\psi}_{\downarrow}\left(\vec{r}'_j\right)\right\rangle \\ &= -(A_{\downarrow\uparrow})_{ji} = -im K_0(m|u_i - w_j|).
\end{aligned}
\end{equation}
\end{subequations}
This trial wavefunction $\Psi_{\mathbb{Z}_{4}}$ can be understood as composite fermions developing intra-species $p$-wave pairing ($p_{x}\pm ip_{y}$ for the two spins) and inter-species $s$-wave pairing. Specifically, $A_{\uparrow\uparrow}$ and $A_{\downarrow\downarrow}$ describe the intra-species $p$-wave pairing wavefunction, while $A_{\uparrow\downarrow}$ describes the inter-spin-species $s$-wave pairing wavefunction.

\para
We now examine the asymptotic behavior of the pairing wavefunction. At small $m|\vec{r}|$, the Bessel function behaves as $K_0(m|\vec{r}|)=-\gamma+\log 2-\frac{1}{2}\log (m^2 |\vec{r}|^2) + {\cal O}(m^2 |\vec{r}|^2)$, where $\gamma\approx0.577$ is Euler's constant. Therefore, in the limit $m\to0$, the correlator becomes 
\begin{equation}
\begin{aligned}   
& \begin{pmatrix}\left\langle \psi_{\uparrow}\left(\vec{r}\right)\psi_{\uparrow}\left(0\right)\right\rangle  & \left\langle \psi_{\uparrow}\left(\vec{r}\right)\bar{\psi}_{\downarrow}\left(0\right)\right\rangle \\
\left\langle \bar{\psi}_{\downarrow}\left(\vec{r}\right)\psi_{\uparrow}\left(0\right)\right\rangle  & \left\langle \bar{\psi}_{\downarrow}\left(\vec{r}\right)\bar{\psi}_{\downarrow}\left(0\right)\right\rangle 
\end{pmatrix}_{m\to0}  \propto \begin{pmatrix}\frac{1}{z} & 0 \\
 0 & \frac{1}{\bar{z}}
\end{pmatrix},
\end{aligned}
\end{equation}
which recovers the result of the decoupled ${\rm Ising} \times \overline{\rm Ising}$ CFT. In other words, the trial wavefunction recovers the Pfaffian$\times \overline{\rm Pfaffian}$ state in the limit $m=0$. At large distances $m|\vec{r}|\gg1$, the asymptotic behavior of the Bessel function is $K_0 (m|\vec{r}|) \sim e^{-m|\vec{r}|}/|\vec{r}|^{7/2}$. This exponential decay physically manifests the pairing between the composite fermions. In particular, the $K_0$ factor in $(A_{\uparrow\downarrow})_{ij} $ and $(A_{\downarrow\uparrow})_{ij} $ represents the bound state between two species of composite fermions, which arises from the anyonic exciton condensation.

\para
The $\mathbb{Z}_4$ FQSH state supports the Abelian anyons ${\sf e}^p {\sf m}^q$ with $p,q=0,1,2,3$, where ${\sf e}$ and ${\sf m}$ have self-bosonic statistics and mutual braiding statistics ${\cal M}_{\sf em} = e^{i\pi/2}$~\cite{jian_minimal_2025}. ${\sf e}$ has electric charge $Q=e/2$ and no spin $S_z$ quantum number, while ${\sf m}$ is a charge-neutral excitation with $S_z = \hbar/4$. Physically, ${\sf e}$ can be identified as a quarter of an inter-species Cooper pair $f_\uparrow^\dagger f^\dagger_\downarrow$, while ${\sf m}$ can be viewed as a quarter of an exciton $f_\uparrow^\dagger f_\downarrow$. The anyon ${\sf e}^2{\sf m}^2$ can be combined with a local fermion to form a neutral fermion excitation with vanishing charge and spin $S_z$. All these anyonic excitations can be generated by inserting extra operators into the correlator $\Psi_{\mathbb{Z}_{4}} = \left\langle f_\uparrow(\vec{r}_1) f_\uparrow(\vec{r}_2) \cdots f_\uparrow(\vec{r}_N) f_\downarrow(\vec{r}_1') f_\downarrow(\vec{r}_2') \cdots f_\downarrow(\vec{r}'_N) \right\rangle$. 
First, the vertex operators $e^{i n_\uparrow \phi_\uparrow} e^{i \bar{n}_\downarrow \bar{\phi}_\downarrow}$ with even integers $n_\uparrow, \bar{n}_\downarrow\in 2\mathbb{Z}$ generate the subset of Abelian anyons with electric charge $Q= \frac{n_\uparrow+\bar{n}_\downarrow}{4} e$ and spin $S_z=\frac{n_\uparrow-\bar{n}_\downarrow}{8} \hbar$. They correspond to ${\sf e}^{{(\UA{n}+\BDA{n})}/{2}}{\sf m}^{({\UA{n}-\BDA{n}})/{2}}$ in the $\mathbb{Z}_4$ topological order. Inserting $\psi_{\uparrow}$ or $\bar{\psi}_{\downarrow}$ yields the neutral fermionic excitation with zero charge and spin. $\psi_{\uparrow}$ and  $\bar{\psi}_{\downarrow}$ are topologically identified with the same fermionic excitation due to the condensation of $\psi_{\uparrow}\bar{\psi}_{\downarrow}$.

\para
The rest of the Abelian anyons in the $\mathbb{Z}_4$ FQSH state must involve the operators $\sigma_\uparrow$ and $\bar{\sigma}_\downarrow$ from the chiral and anti-chiral Ising CFTs, which require more careful treatments. The operator $\sigma_\uparrow$ ($\bar{\sigma}_\downarrow$) inserts a $\pi$-flux for the (anti-)chiral Majorana fermion $\psi_{\uparrow}$ ($\bar{\psi}_{\downarrow}$). When $\psi_{\uparrow}\bar{\psi}_{\downarrow}$ is condensed, $\sigma_\uparrow$ and $\bar{\sigma}_\downarrow$ must appear together. This is because the last term of the coupled Ising theory Eq.~\eqref{eq:L_coupled_Ising} hybridizes $\UA{\psi}$ and $\BDA{\psi}$. When a $\pi$-flux is introduced, it should now act simultaneously on the Majorana fermion fields of both species (or chirality) to avoid explicit branch cuts in the wavefunction. In the 2D Ising CFT (that includes both the chiral and anti-chiral sectors), it is known that the product of the primary fields $\sigma_\uparrow (z) \bar{\sigma}_\downarrow(\bar{z})$ can be alternatively captured by the spin operator $s(z,\bar{z})$ and disordered operator $\mu (z,\bar{z})$~\cite{DiFrancescoCFT}. The remaining Abelian anyons of the $\mathbb{Z}_4$ FQSH state can be inserted using the operators $s \, e^{i n_\uparrow'\phi_\uparrow}e^{i \bar{n}_\downarrow'\bar{\phi}_\downarrow} $ and $\mu \, e^{i n_\uparrow'\phi_\uparrow}e^{i \bar{n}_\downarrow'\bar{\phi}_\downarrow} $ with odd integers $\UA{n}', \BDA{n}'\in 2\mathbb{Z}+1$. In particular, we identify $s e^{i \phi_\uparrow}e^{i \bar{\phi}_\downarrow} \sim {\sf e}$ and $s e^{i \phi_\uparrow}e^{-i \bar{\phi}_\downarrow} \sim {\sf m}$. One can readily show that this identification is consistent with the electric charge and the $S_z$ quantum number of the anyons ${\sf e}$ and ${\sf m}$.

\para
Furthermore, we can verify that the operators $s \, e^{i n_\uparrow'\phi_\uparrow}e^{i \bar{n}_\downarrow'\bar{\phi}_\downarrow} $ and $\mu \, e^{i n_\uparrow'\phi_\uparrow}e^{i \bar{n}_\downarrow'\bar{\phi}_\downarrow} $ are compatible with the fusion rules and Abelian statistics of the $\mathbb{Z}_4$ topological order. For the Ising CFT in the decoupled limit ($m=0$), the operators $s$ and $\mu$ satisfy the fusion rules (extracted from their operator product expansions)~\cite{DiFrancescoCFT}
\begin{align} \label{eq:Ising_Fusion}
    s\times s \rightarrow 1+ \psi_\uparrow \bar{\psi}_\downarrow,~~~ &\mu\times \mu \rightarrow 1+ \psi_\uparrow \bar{\psi}_\downarrow, \nonumber \\
    s\times \psi_\uparrow \rightarrow \mu,~~~& s\times \bar{\psi}_\downarrow \rightarrow \mu, \\
    \mu\times \psi_\uparrow \rightarrow s,~~~& \mu\times \bar{\psi}_\downarrow \rightarrow s. \nonumber  
\end{align} 
In the $\mathbb{Z}_4$ FQSH state, $\psi_{\uparrow}\bar{\psi}_{\downarrow}$ should be identified with the identity operator due to its condensation when $m\neq 0$. Consequently, the fields $s$ and $\mu$ obey the $\mathbb{Z}_2$ fusion rules in the coupled Ising theory Eq.~\eqref{eq:L_coupled_Ising}. Considering that $\UA{\psi} \BDA{\psi} e^{i 4\phi_\uparrow}  e^{i 4\bar{\phi}_\downarrow}$ corresponds to a local inter-species Cooper pair, which is a topologically trivial excitation, we can confirm that the anyon $s e^{i \phi_\uparrow}e^{i \bar{\phi}_\downarrow} \sim {\sf e}$ indeed follows the $\mathbb{Z}_4$ fusion rule and can be viewed as a quarter of the inter-species Cooper pair. The $\mathbb{Z}_4$ fusion rule of $s e^{i \phi_\uparrow}e^{-i \bar{\phi}_\downarrow} \sim {\sf m}$ can be verified in a similar manner. The braiding statistics of $s e^{i \phi_\uparrow}e^{i \bar{\phi}_\downarrow}$ and $s e^{i \phi_\uparrow}e^{-i \bar{\phi}_\downarrow}$ solely come from the bosonic sectors and agree with the expectations for the anyons ${\sf e}$ and ${\sf m}$. Notice that the operator $s$ contributes trivially to the statistics as it descends from the conjugate pair $\UA{\sigma} \BDA{\sigma}$. The anyons associated with $\mu e^{i \phi_\uparrow}e^{i \bar{\phi}_\downarrow}$ can be viewed as those generated by combining $s e^{i \phi_\uparrow}e^{i \bar{\phi}_\downarrow}$ with an extra neutral fermion $\UA{\psi}$ or $\BDA{\psi}$. 

\para
We remark that, before the anyonic exciton condensation, anyons that involve $\sigma_\uparrow (z) \bar{\sigma}_\downarrow(\bar{z})$ are non-Abelian excitations of the Pfaffian$\times \overline{\rm Pfaffian}$ state. The condensation of $\UA{\psi}\BDA{\psi}$ causes $\sigma_\uparrow (z) \bar{\sigma}_\downarrow(\bar{z})$ to split into $s(z,\bar{z})$ and $\mu(z,\bar{z})$. Additionally, as mentioned before, isolated $\sigma_\uparrow (z)$ or $ \bar{\sigma}_\downarrow(\bar{z})$ are disallowed to avoid explicit branch cuts in the wavefunction. As a consequence, there is no non-Abelian excitation in the $\mathbb{Z}_4$ FQSH state. This result is consistent with the purely topological-quantum-field-theoretic analysis of the anyonic exciton condensation discussed in Ref.~\cite{jian_minimal_2025}. Furthermore, the result here provides a recipe to construct wavefunctions with anyon excitations in the $\mathbb{Z}_4$ FQSH state.

\section{Monte Carlo analysis of energetics}

\para
To assess the viability of our trial wavefunction $\Psi_{\mathbb{Z}_{4}}$, we use Monte Carlo sampling to estimate its energetics.
To avoid boundary effects and preserve rotational invariance, we evaluate the wavefunctions on a spherical geometry~\cite{haldane_fractional_1983}.
For states in a single LL, the standard approach is to map the two-dimensional planar coordinates to the surface of a sphere enclosing a magnetic monopole, ensuring a uniform perpendicular magnetic field. For our system, we consider the two spin species of fermions experiencing opposite magnetic fields on the sphere. Hence, the single-particle kinetic energy can be expressed as $\hat{T} = \frac{1}{2m_e}(\hat{\vec{p}} - \tau \vec{A})^2$, where $\tau=\pm$ for the two spins, $\vec{A}$ is the vector potential, and $m_e$ is the electron's mass. We assume a two-body power-law decaying (Coulomb) interaction:
\begin{equation}\label{eq:interaction_power_law}
V(r,\tau_1,\tau_2) = 
            \begin{cases} 
            V_{\text{intra}}\left(\frac{\ell_{B}}{r}\right), & \tau_{1}=\tau_{2} \\ 
            V_{\text{inter}}\left(\frac{\ell_{B}}{r}\right), & \tau_{1}\neq\tau_{2} 
            \end{cases}
\end{equation}
where $\ell_{B}$ is the magnetic length and $r$ is the chord distance on the sphere. 
The parameters $V_{\text{intra}}$ and $V_{\text{inter}}$ allow us to tune the intra- and inter-spin-species interaction strength independently. In the following, we will use Monte Carlo sampling to numerically evaluate the energy expectation value of the $\mathbb{Z}_4$ FQSH state $\Psi_{\mathbb{Z}_{4}}$ and compare with other candidate states for the half-filled conjugate LL.

\para
In our Monte Carlo calculation, we estimate the total energy $\langle \hat{H} \rangle = \langle \hat{T}_{\rm total} \rangle + \langle \hat{V} \rangle$ by importance-sampling many-body configurations from $|\Psi|^2$. Here, $\hat{T}_{\rm total} $ and $\hat{V}$ denote the total many-body kinetic and interaction energies. 
The interaction energy $\langle \hat{V} \rangle$ is obtained by summing all pairwise two-body contributions for each sampled configuration.
A crucial feature of $\Psi_{\mathbb{Z}_{4}}$ is that for any finite $m\neq 0$ it does not lie entirely within the lowest Landau level (LLL), so its kinetic energy does not vanish.
Rather than applying a LLL projection ${\cal P}_{\rm LLL}$, which would greatly complicate the wavefunction's analytic structure, we directly evaluate $\langle\hat{T}_{\text{total}}\rangle = \sum_{i} \langle\hat{T}_{i}\rangle = \frac{1}{2m_e} \sum_{i} \langle(\hat{\vec{p}_{i}} - \tau_{i} \vec{A})^2\rangle$ as a sum of single-particle contributions involving derivatives of the many-body wavefunction.
Full details of the sampling procedure are given in Appendix~\ref{app:monte_carlo_sampling}.

\begin{figure}
    \centering
    \includegraphics[width=\linewidth]{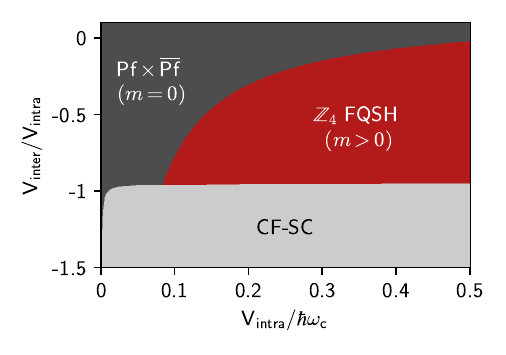}
    \caption{Phase diagram obtained by variational Monte Carlo for the different candidate states in two conjugate half-filled Landau levels. By tuning the interaction-to-kinetic-energy scale $V_{\text{intra}}/\hbar\omega_{c}$ and the ratio of inter- to intra-spin-species interaction strength $V_{\rm inter}/V_{\rm intra}$, we find that all three phases are present: the decoupled ${\rm Pfaffian}\times\overline{\rm Pfaffian}$ state ($m=0$) at weak inter-spin-species interaction or large kinetic energy; the new $\mathbb{Z}_4$ fractional quantum spin Hall state ($m>0$) at stronger inter-spin-species attraction and weaker kinetic energy; and the CF-SC state (pure $s$-wave inter-species BCS state of composite fermions) at strong enough inter-spin-species attraction.}
    \label{fig:mc_phase_diagram}
\end{figure}

\para
We perform Monte Carlo simulations for $N=10$ fermions per spin species (20 in total), beyond the reach of existing exact diagonalization on relevant systems~\cite{kwan_regarding_2026}. For each $V_\text{intra}$ and $V_\text{inter}$, we vary the coupling parameter $m$ to search for the lowest-energy state.  A finite $m$ causes the wavefunction to partially deviate from the LLL. Therefore, the kinetic energy increases monotonically with $m$. If the interactions are purely repulsive, i.e., $V_{\text{intra}}>0$  and $V_{\text{inter}}>0$, our simulation shows that the decoupled ${\rm Pfaffian}\times\overline{\rm Pfaffian}$ state ($m=0$) always remains energetically favorable over the trial $\mathbb{Z}_4$ FQSH states with $m\neq0$.
However, if we maintain a repulsive intra-species interaction ($V_{\text{intra}}>0$) but introduce an attractive inter-species interaction ($V_{\text{inter}}<0$), the optimal coupling $m$ can become nonzero. 
This optimal value is determined by a competition between the interaction energy, which is lowered by increasing $m$ (stronger excitonic binding), and the kinetic energy, which penalizes larger $m$.
We demonstrate this competition in Fig.~\ref{fig:mc_phase_diagram}, which maps the phase diagram as a function of the interaction-to-kinetic-energy scale $V_{\text{intra}}/\hbar\omega_{c}$ and the relative inter-spin-species interaction strength $V_{\rm inter}/V_{\rm intra}$.
We find that the coupled, finite-$m$ state is favored when the inter-species attraction is sufficiently strong, and the kinetic energy penalty does not entirely dominate. We remark that a bare attractive interaction between electrons of different spins is unlikely in realistic materials. However, when we replace the electrons in one of the spin species by holes, an attractive inter-spin-species interaction becomes natural. An example of such a scenario is the TMD heterobilayer~\cite{nguyen_quantum_2025,chang_electric-field-tuned_2026}, where the layer index serves as the pseudospin.

\para
In the regime with attractive $V_{\rm inter}<0$, the $\mathbb{Z}_4$ FQSH state $\Psi_{\mathbb{Z}_{4}}$ may compete with another trial state given by the pure inter-species $s$-wave superconductor of composite fermions (CF-SC), which has been previously studied in Refs.~\cite{wagner_variational_2025,wagner_s-wave_2021} in slightly different settings. This trial wavefunction of a composite-fermion $s$-wave superconductor takes the form
\begin{equation}
\begin{aligned}    
\Psi_\text{CF-SC}&\left( \left\{ u_{i} \right\}, \left\{ w_{i} \right\} \right) = \prod_{i<j}(u_{i}-u_{j})^{2} \prod_{i<j}(\bar{w}_{i}-\bar{w}_{j})^{2} \\
&\times \exp\left[-\frac{1}{4\ell_{B}^2}\sum_i \left(|u_i|^2 + |w_i|^2 \right) \right]
 {\rm det}(G).
\end{aligned}
\end{equation}
Here $G$ is an $N\times N$ matrix with elements
\begin{equation}
  G_{ij} = \sum_{n\geq 0}\sum_{k}\, e^{\alpha n}\, \phi_{n,k}(u_i)\,\overline{\phi}_{n,k}(w_j),
\end{equation}
where $n = 0,1,2,\ldots$ labels the LLs for a given spin species ($n=0$ is the LLL), $k$ labels the degenerate orbitals within the $n$th LL, and $\phi_{n,k}$ is the corresponding normalized single-particle eigenstate on the sphere.
The exponential weight $e^{\alpha n}$ suppresses contributions from higher LLs for $\alpha<0$: in the limit $\alpha\to-\infty$ only the LLL term survives, so $G$ reduces to the LLL projector and $\Psi_{\rm CF\text{-}SC}$ becomes fully LLL-projected. Finite $\alpha$ allows higher-LL mixtures and is optimized variationally to minimize the total energy. 
When the fermions of both spin species are electrons, the state $\Psi_\text{CF-SC}$ is topologically equivalent to an $s$-wave BCS superconductor. If the fermions in one of the species are holes instead, $\Psi_\text{CF-SC}$ describes an exciton insulator without topological order.

\para
Using identical Monte Carlo sampling techniques, we estimate the kinetic and interaction energies of $\Psi_\text{CF-SC}$ and optimize $\alpha$ to minimize the total energy.
Unlike Refs.~\cite{wagner_variational_2025,wagner_s-wave_2021}, which explicitly apply ${\cal P}_{\rm LLL}$ to $\Psi_\text{CF-SC}$, we treat the kinetic energy explicitly for comparison with the FQSH state $\Psi_{\mathbb{Z}_{4}}$. The resulting phase boundaries based on the Monte Carlo energy estimates are depicted in Fig.~\ref{fig:mc_phase_diagram}, highlighting the parameter regimes where each wavefunction minimizes the ground state energy.
Ultimately, we find that our proposed trial wavefunction $\Psi_{\mathbb{Z}_{4}}$ has a favorable energy expectation value (at least within the states we considered) in regimes where the attractive inter-species interaction is strong enough to overcome the intra-species repulsion and the kinetic energy cost, and stabilize the anyonic exciton condensate. In principle, one can also perform the projection ${\cal P}_{\rm LLL}$ to both $\Psi_\text{CF-SC}$ and $\Psi_{\mathbb{Z}_4}$ and compare the interaction energies of these projected states. We will leave this investigation for future exploration. Also, we remark that our analysis on the specific trial wavefunction $\Psi_{\mathbb{Z}_{4}}$ here does not necessarily rule out the potential emergence of the $\mathbb{Z}_4$ topological order under purely repulsive interactions. Constructing alternative FQSH trial wavefunctions with $\mathbb{Z}_4$ topological order that are not penalized by the inter-species repulsion is an interesting direction for future research.

\section{Summary and outlook}

\para
We have constructed and evaluated a variational wavefunction for a $\mathbb{Z}_4$ fractional quantum spin Hall state in a pair of conjugate Landau levels half-filled with fermions with two (pseudo) spin species. The construction couples the chiral and anti-chiral Ising sectors of a pair of Pfaffian and conjugate Pfaffian states through a variational mass parameter, effectively describing the condensation of an anyonic exciton. This $\mathbb{Z}_4$ state provides a realization of the recently proposed \emph{minimal} FQSH topological order~\cite{jian_minimal_2025} compatible with the transport signatures observed in recent FQSH experiments~\cite{kang_evidence_2024}; our construction explicitly realizes this minimal topological order as a concrete microscopic trial wavefunction. Using Monte Carlo simulations for energy estimations, we mapped a variational phase diagram (Fig.~\ref{fig:mc_phase_diagram}) comparing the $\mathbb{Z}_4$ FQSH state with the decoupled ${\rm Pfaffian}\times\overline{\rm Pfaffian}$ pair and the composite-fermion $s$-wave superconductor. The $\mathbb{Z}_4$ state becomes energetically favorable when an attractive inter-spin-species interaction is strong enough to overcome the kinetic energy penalty of LL mixing. Such an attraction arises naturally in the electron-hole setting, as in TMD heterobilayers~\cite{nguyen_quantum_2025,chang_electric-field-tuned_2026}. Concrete material platforms will be described by different Hamiltonians than the simple one we used, which will inevitably deform the phase boundaries of Fig.~\ref{fig:mc_phase_diagram}, but our results show that the $\mathbb{Z}_4$ state can occupy a finite region of phase space.

\para
Looking forward, there are several natural directions for future research. First, it will be valuable to benchmark our variational phase diagram against exact diagonalization studies of microscopic models~\cite{kwan_regarding_2026}. Such numerical studies, particularly those employing realistic effective interactions for bilayer TMDs, will be vital to examine whether specific moir\'{e} or heterostructure platforms naturally reside within the stability window of the $\mathbb{Z}_4$ state. In particular, the conjugate Chern bands formed by electrons, such as twisted TMD homobilayers~\cite{kang_evidence_2024,kang_time-reversal_2025,wang_magnetic_2026}, and those that consist of both electrons and holes, such as TMD heterostructures~\cite{nguyen_quantum_2025,chang_electric-field-tuned_2026}, have drastically different microscopic Hamiltonians for this examination. Moreover, the conjugate Chern bands in these systems generally exhibit non-ideal quantum geometries. Investigating the possible realization of the $\mathbb{Z}_4$ fractional quantum spin Hall state in these realistic material platforms will be an interesting future direction. Furthermore, the theoretical mechanism employed here---condensing anyonic excitons formed by the neutral fermionic excitations of decoupled conjugate topological states---is highly versatile. We anticipate that this general recipe can be extended well beyond the $\text{Pfaffian}\times\overline{\text{Pfaffian}}$ starting point to systematically construct and evaluate new families of fractional quantum spin Hall trial wavefunctions.

\section*{Acknowledgement}
We thank M.~Yutushui for useful discussions.
O.L. is supported by a Bethe-KIC postdoctoral fellowship at Cornell University. C.-M.J. is supported by the Alfred P. Sloan
Foundation through a Sloan Research Fellowship.

\appendix


\section{Details of Monte Carlo simulations}\label{app:monte_carlo_sampling}

\subsection{Sampling and interaction energy}
To avoid boundary effects, we evaluate the wavefunctions on a Haldane sphere~\cite{haldane_fractional_1983} of radius $R = \sqrt{Q} \ell_{B}$, which encloses a magnetic monopole of strength $2Q$. A particle position is parameterized by the polar and azimuthal angles, $\mathbf{r}_i = (\theta_i, \phi_i)$. Configurations $\mathbf{R} = \{\mathbf{r}_1, \ldots, \mathbf{r}_N\}$ are sampled according to the probability distribution $P(\mathbf{R}) \propto |\Psi_{\rm T}(\mathbf{R})|^2 \prod_i \sin(\theta_i)$ using the Metropolis algorithm. The inclusion of the $\sin(\theta_i)$ factor ensures the correct spherical integration measure. The procedure is as follows:
\begin{enumerate}
\item Propose a new configuration by randomly displacing a single particle $i$: $\theta_{i,\text{new}} = \theta_{i,\text{old}} + \delta\theta$ and $\phi_{i,\text{new}} = \phi_{i,\text{old}} + \delta\phi$.
\item Calculate the acceptance probability, accounting for the spherical measure:
\begin{equation}
p_{\text{acc}} = \min\left(1, \frac{|\Psi_{\rm T}(\mathbf{R}_{\text{new}})|^2 \sin(\theta_{i,\text{new}})}{|\Psi_{\rm T}(\mathbf{R}_{\text{old}})|^2 \sin(\theta_{i,\text{old}})}\right).
\label{eq:metropolis}
\end{equation}
\item Accept or reject based on a random number comparison. Moves that push a particle outside the polar bounds $\theta \in [0, \pi]$ are strictly rejected.
\end{enumerate}

The interaction energy is computed by evaluating the local energy $E_{\rm L}(\mathbf{R}) = \hat{V}\Psi_{\rm T}(\mathbf{R}) / \Psi_{\rm T}(\mathbf{R})$. For the spherical geometry, the interaction potential is evaluated using the 3D chord distance between particles,
\begin{equation}
d_{\text{chord}}(\mathbf{r}_i, \mathbf{r}_j) = \sqrt{(x_i - x_j)^2 + (y_i - y_j)^2 + (z_i - z_j)^2},
\end{equation}
where $x, y, z$ are the standard Cartesian coordinates of the particles on the unit sphere. The total interaction energy is then estimated as the sample average $\langle \hat{V} \rangle \approx \frac{1}{N_{\text{samples}}} \sum_{k=1}^{N_{\text{samples}}} E_{\rm L}(\mathbf{R}_k)$.

\subsection{Kinetic energy}
We now discuss the evaluation of the kinetic energy, which is non-zero because the trial wavefunction $\Psi_{\mathbb{Z}_{4}}$ does not reside entirely in the lowest Landau level for finite coupling $m\neq0$. 
On the spherical geometry, the kinetic energy operator $\hat{T}$ is proportional to the squared dynamical angular momentum operator $\mathbf{\Lambda}^2$~\cite{haldane_fractional_1983,jain_landau_2007}. For a particle with charge $q_i$ subjected to the uniform radial magnetic field generated by the central monopole, the kinetic energy operator in the symmetric gauge (with Dirac strings traversing through the poles) acts on the angular coordinates as:
\begin{equation}\label{eq:spherical_kinetic}
\begin{aligned}
\hat{T}_i &= -\frac{\partial^2}{\partial \theta_i^2} - \cot\theta_i \frac{\partial}{\partial \theta_i} - \frac{1}{\sin^2\theta_i} \frac{\partial^2}{\partial \phi_i^2} \\
&\quad + (q_i \cot\theta_i)^2 + \frac{2iq_i \cot\theta_i}{\sin\theta_i} \frac{\partial}{\partial \phi_i},
\end{aligned}
\end{equation}
where $q_i = Q$ for electrons and $q_i = -Q$ for holes. 

For states residing strictly within the lowest Landau level, the expectation value of $\hat{T}_i$ for each particle is exactly equal to the zero-point energy $Q$. To isolate the kinetic energy penalty arising from LL-mixing, we subtract this baseline zero-point energy, defining the total effective kinetic energy operator as $\hat{T}_{\text{total}} = \sum_i \hat{T}_i - N Q$.

To evaluate the expectation value $\langle \hat{T}_{\text{total}} \rangle$ using Monte Carlo methods, we employ the local energy approach. At each accepted configuration $\mathbf{R}$, we compute the local kinetic energy:
\begin{equation}
E_{\text{loc}}(\mathbf{R}) = \frac{\hat{T}_{\text{total}}\Psi_{\rm T}(\mathbf{R})}{\Psi_{\rm T}(\mathbf{R})} = \sum_{i=1}^N \frac{\hat{T}_i \Psi_{\rm T}(\mathbf{R})}{\Psi_{\rm T}(\mathbf{R})} - NQ.
\end{equation}
The derivatives in Eq.~\eqref{eq:spherical_kinetic} are computed numerically via central finite differences. For each particle $i$, we evaluate the many-body wavefunction at displaced coordinates $(\theta_i \pm \delta, \phi_i)$ and $(\theta_i, \phi_i \pm \delta)$, where $\delta \approx 10^{-4}$ is a small step size. The first and second derivatives are then approximated as:
\begin{equation}
\frac{\partial \Psi_{\rm T}}{\partial \theta_i} \approx \frac{\Psi_{\rm T}(\theta_i + \delta) - \Psi_{\rm T}(\theta_i - \delta)}{2\delta},
\end{equation}
\begin{equation}
\frac{\partial^2 \Psi_{\rm T}}{\partial \theta_i^2} \approx \frac{\Psi_{\rm T}(\theta_i + \delta) - 2\Psi_{\rm T}(\theta_i) + \Psi_{\rm T}(\theta_i - \delta)}{\delta^2},
\end{equation}
with identical expressions for the azimuthal derivatives with respect to $\phi_i$. The expectation value of the kinetic energy is then obtained by averaging $E_{\text{loc}}(\mathbf{R})$ over all uncorrelated Monte Carlo samples.

\bibliography{library}

\end{document}